\documentclass[journal]{IEEEtran}
\usepackage{amsmath,amssymb,amsthm,booktabs,graphicx,cite,xcolor,algorithm,algpseudocode}
\newtheorem{proposition}{Proposition}
\newtheorem{lemma}{Lemma}
\newcommand{\R}{\operatorname{Re}}
\newcommand{\I}{\operatorname{Im}}
\newcommand{\E}{\mathbb{E}}

\begin{document}
\title{Nearest Is Not Safest: Fluid-Antenna Port Selection\\for Over-the-Air Backpropagation in Urban Multipath}
\author{Xianzhong~Li, Yuzhi~Yang, Tianhao~Guo
\thanks{Xianzhong Li and Tianhao Guo are with the College of Physics and Electronic Engineering, Shanxi University, Taiyuan 030006, China (e-mail: \{lixianzhong, tianhao\_guo\}@sxu.edu.cn).}%
\thanks{Yuzhi Yang is with the College of Information Science, Japan Advanced Institute of Science and Technology, Nomi 9231292, Japan (e-mail: yangyzis@jaist.ac.jp).}%
\thanks{Corresponding author: Tianhao Guo.}}
\maketitle

\begin{abstract}
Over-the-air backpropagation returns a loss gradient through the reciprocal channel, which requires the return port to see the channel of the forward port. If the terminal moves in between, a fluid antenna system can return through the port nearest to the forward position. In ray-traced Chicago streets, with a pedestrian displacement of about half a wavelength and a street-level base station, this rule reverses 20 percent of the gradients at a port spacing of 0.7 wavelengths and 49 percent at 0.9 wavelengths, while one of the four nearest ports stays aligned in all but 2 and 9 percent. We derive a certificate for the sign of the mean correlation from the second moment of a planar angular spectrum. It reduces to a linear program with three variables and infinitely many constraints, is tight, and holds over a region of second moments. In a static Gaussian model, observations collected along a line cannot tell an angular spectrum from its mirror image, so the sign of the cross moment must come from other information. At unsurveyed positions of the same city, a knowledge map that transports surveyed paths through virtual sources and fits them to the causal channel history lowers the reversal rate at 0.9 wavelengths from 49 to 27 percent. Its gain over the nearest port survived every tested error, while its gain over a geometry rule stayed significant at the tested single-error levels of 5 degrees of phase drift per step and 2 mm of position error but not at the next tested levels. Without any survey, across five cities and ten base stations, the nearest port reverses 37 percent of the returns at 0.9 wavelengths and a rule based on the direction and elevation of the base station 20.5 percent. The multilayer perceptrons we tested did not beat these physical rules by the margin fixed in advance in the primary settings.
\end{abstract}


\section{Introduction}
Over-the-air split learning computes the backward pass of a wireless layer by sending the receiver's loss gradient back through the channel. Reciprocity then applies the adjoint of the forward channel, so the transmitter needs no channel estimate and no high-dimensional gradient is fed back digitally~\cite{yang2023otasl}. We consider the backward transmission of a single wireless layer, in which the terminal equalizes the forward channel and returns a scalar gradient signal. For such a scalar layer a digital return of the loss derivative would also be cheap, and we use it to isolate the alignment question that every output of a wireless layer faces. The condition is that the return sees the forward channel. A moving terminal breaks this condition, and the returned gradient is attenuated, rotated, or reversed~\cite{truong2013aging,li2026coherent}.

Two remedies keep the return channel close to the forward one. A movable element can return to its forward world position, and a terminal with many fixed ports can switch the return to the port whose world position is nearest to the forward position. The second option suits a fluid antenna system, whose ports are densely packed and selected electronically~\cite{wong2021fas}. Below a residual displacement of a quarter wavelength, the mean real correlation between the two ports stays positive for every angular spectrum, which makes the mean returned gradient a descent direction under the scalar conditional-mean model of~\cite{li2026keep}. A positive mean correlation does not prevent the reversal of an individual return, so we report the empirical reversal rate separately from any statement about the mean.

Street canyons are not isotropic. Energy arrives from a few directions, and the correlation of the channel falls much faster along some directions than along others. A port that is nearer in Euclidean distance can then be farther in the metric that matters. We measure how often this happens in ray-traced cities, characterize the information that a safe choice requires, and test whether that information can be obtained at positions that were never surveyed. No rule in this paper measures the candidate return channels at the return time. The deployable rules use, in increasing order, the positions and the heading of the terminal, a survey of path statistics at other positions in the same city, and the causal history of the forward channel, while two reference rules use the paths or the channels at the query position.

\emph{Contributions.}
\begin{itemize}
\item \emph{A certificate from second moments.} For any normalized planar angular spectrum with a given second-moment matrix, the worst and best mean real correlation at a residual follow from a linear program with three variables and infinitely many constraints. One dual solution covers a region of second moments and a ball of position errors, which gives closed-form linear tolerances (Section~\ref{sec:cert}).
\item \emph{The information a safe choice needs.} The second moment does not rank ports in general. In synthetic two-cluster spectra the tested second-moment rules leave a gap that the fourth moments close, whereas in the 3GPP CDL profiles they are nearly sufficient. Under a static Gaussian model, observations collected along a line cannot distinguish a spectrum from its mirror image, so the sign of the cross moment must come from other information, such as a map or the direction of the base station (Section~\ref{sec:stats}, Appendices~\ref{app:synth} and~\ref{app:mirror}).
\item \emph{Ray-traced evidence.} In the Chicago scene of an open channel-generation pipeline, with pedestrian motion and a finite fluid-antenna aperture, nearest-port selection reverses the returned gradient in up to 49 percent of the exchanges for the street-level base station and in at most 5.8 percent for a rooftop base station of the same scene at the spacings of Table~\ref{tab:real4}. In four further cities the nearest port reverses between 26 and 54 percent of the returns at 0.9 wavelengths, including rooftop sites (Section~\ref{sec:eval}, Appendices~\ref{app:veh} and~\ref{app:cities}).
\item \emph{Knowledge maps for unsurveyed positions and their error budget.} In the zero-scattering street-level data, a map that transports surveyed paths through their virtual sources and fits path amplitudes to the causal history of the forward channel lowers the reversal rate from 0.32 to 0.27 at 0.9 wavelengths. An error budget over history noise, phase drift, position, orientation and calibration errors shows which of its gains survive (Sections~\ref{sec:ckm} and~\ref{sec:budget}, Appendices~\ref{app:real3} and~\ref{app:budget}).
\item \emph{Effect on training.} On a scalar link with a power-limited return, link noise and channel estimation error, the nearest port lowers the test accuracy of a small classifier by about 13 to 15 points, and the cosine between the returned and the exact gradient orders the rules exactly as their reversal rates do (Section~\ref{sec:train}, Appendix~\ref{app:train}).
\end{itemize}

\section{Related Work}
\emph{Over-the-air learning and channel aging.} Analog over-the-air computation aggregates model updates by the superposition of the channel~\cite{zhu2020analog,yang2020otafl}, and over-the-air split learning obtains the backward pass of a wireless layer by reciprocity~\cite{yang2023otasl}. Training physical systems through their own response is also pursued outside wireless~\cite{wright2022pnn}, and model-free training of communication links replaces the channel gradient by a loss feedback~\cite{aoudia2019modelfree}. Channel aging between the forward and backward transmissions has been studied for massive MIMO precoding~\cite{truong2013aging} and countered by angular-delay prediction~\cite{yin2020prony}. Its effect on the returned gradient is a rotation set by the phase of the temporal correlation~\cite{li2026coherent}, which can be waited out at a single antenna~\cite{li2026wait}. Restoring the forward position mechanically or by port relabeling keeps the mean correlation positive for every spectrum below a quarter-wavelength residual~\cite{li2026keep}, and assigning the ports of an array to the displaced ports with the help of angular statistics recovers the aged return in its simulations~\cite{li2026coherent}. That assignment is made for all ports of an array at once. The choice of a single return port from path-level information in ray-traced streets with a finite aperture is not addressed there.

\emph{Fluid antenna systems.} Fluid antenna systems select among densely packed ports electronically and model the port correlation by the spatial correlation of the field~\cite{wong2020limits,wong2021fas}. Port selection serves multiple access~\cite{wong2022fama}, and under mobility a predicted port can keep the channel close to a reference channel~\cite{fasmobility2024}. These selections target the received power, the rate, or the deviation from a reference channel. Under the unit-power model of Section~\ref{sec:model} the expected squared deviation of a port from the forward channel equals $2-2c(\boldsymbol r)$, so choosing the port with the largest mean real correlation is the same as choosing the port with the smallest expected deviation. The two statistical objectives therefore coincide in the mean. For a single return they differ, because the squared deviation also contains the power of the return channel, and the probability that the update is reversed depends on the phase of the correlation (Lemma~\ref{lem:risk}). Our contribution lies in the certificate on the sign of the mean from limited information, in the limits of what collinear observations reveal, in the reversal rates of the nearest port, and in the information levels at which a better port can be chosen.

\emph{Channel knowledge maps and digital twins.} A channel knowledge map stores location-specific channel information and interpolates it to unsurveyed positions~\cite{zeng2021ckm,zeng2024ckm}. Learned radio maps~\cite{levie2021radiounet} and channel charts~\cite{studer2018charting} infer such information from data. Ray tracing on a city model provides the same information at any position when the geometry and materials are known. We use ray tracing as the ground truth and compare knowledge maps built from a survey of the same scene.

\section{System Model}\label{sec:model}
\subsection{Over-the-Air Backpropagation}
The transmitter sends a real scalar symbol $q=\boldsymbol w^{\mathsf T}\boldsymbol x$ with real parameters $\boldsymbol w$ and real features $\boldsymbol x$, and the terminal receives $y=h_fq+n$ with a nonzero forward channel $h_f$. The terminal equalizes with an accurate estimate of $h_f$, computes a real loss of the equalized output, and returns $\delta=2\,\partial\mathcal L/\partial y^*$ at time $t_r=t_f+\tau$ through the channel $h_r$ of the return port. The terminal sends $\delta^*$, the reciprocal channel multiplies it by $h_r$, and after reciprocity calibration and conjugation of the received signal the transmitter obtains $\R(h_r^*\delta)\boldsymbol x$ plus noise, whereas the exact gradient is $\R(h_f^*\delta)\boldsymbol x$. In this model the returned gradient equals the exact one multiplied by
\begin{equation}\label{eq:s}
s=\frac{\R(h_r h_f^*)}{|h_f|^2},
\end{equation}
plus noise (Appendix~\ref{app:s}). A practical return scales $\delta$ to a power budget, which multiplies the update by a positive factor and leaves the sign of $s$ unchanged. Under the nondegenerate jointly Gaussian model of Section~\ref{sec:stats}, with $|\phi|<1$, $s$ has infinite variance, which is why the training surrogate of Section~\ref{sec:train} clips it. For other receivers and multi-output layers, $s$ is a proxy for the alignment of the return. The terminal uses its forward channel estimate for equalization but needs no estimate of the return channel. The noiseless part of the update has a positive projection on the exact gradient when $s>0$ and a negative one when $s<0$. We call the event $s<0$ a \emph{reversal}.

\subsection{Fluid-Antenna Ports and Motion}
The terminal carries a fluid antenna system with ports on a triangular lattice of spacing $d$ inside a disc of radius $1.5\lambda$ in the body frame. The forward transmission uses the port at the body origin. Between the two transmissions the body moves by $\boldsymbol D=\boldsymbol v\tau$. A return port at body position $\boldsymbol p$ then lies at the world residual $\boldsymbol r=\boldsymbol p+\boldsymbol D$ from the forward position, and the four ports with the smallest $\|\boldsymbol r\|$ are the candidates. The aperture holds 37, 19, 7 and 7 ports for spacings of $0.5\lambda$, $0.7\lambda$, $0.9\lambda$ and $1.2\lambda$, counting the six ports on the boundary at $0.5\lambda$, so the spacing sweep changes the number of ports together with their density. We fix four candidates to keep the search local. This preselection by distance can exclude a better port, and we do not quantify this loss. The reference rules Orc and Spec below choose among these four. The selection rule must pick one of them without a channel estimate at the return time. When $\|\boldsymbol D\|+d/\sqrt3\le1.5\lambda$, the lattice point nearest to the forward position lies in the aperture and the nearest candidate has $\|\boldsymbol r\|\le d/\sqrt3$. All pedestrian settings below satisfy this condition.

\subsection{Correlation Model}
Let $\mu$ be the angular power measure at the forward position, normalized to unit mass, and $\boldsymbol u_h$ the horizontal projection of the arrival direction. Under uncorrelated scattering the mean correlation between the forward port and a port at residual $\boldsymbol r$ is
\begin{equation}\label{eq:phi}
\phi(\boldsymbol r)=\int e^{jk\boldsymbol u_h^{\mathsf T}\boldsymbol r}\,\mu(d\boldsymbol u),\qquad c(\boldsymbol r)=\R\phi(\boldsymbol r),
\end{equation}
with $k=2\pi/\lambda$. The arrival direction points from the receiver to the source, so a move toward the source raises the phase. The horizontal projection has length $\sin\theta$ for zenith angle $\theta$ and is not renormalized. The quantity $c(\boldsymbol r)$ is the mean of $\R(h_rh_f^*)$ for unit-power channels, which we call the mean real correlation. Under general channel laws $c$ need not equal the mean of the normalized factor $s$, and the two coincide under the Gaussian model of Lemma~\ref{lem:risk}. A single realization can have $s<0$ when $c>0$.

\section{A Certificate from Second Moments}\label{sec:cert}
A terminal rarely knows $\mu$. It may know, or estimate, the horizontal second moment $C=\int\boldsymbol u_h\boldsymbol u_h^{\mathsf T}d\mu$. For a planar spectrum with $\boldsymbol u_h=(\cos\vartheta,\sin\vartheta)$, knowledge of $C$ is knowledge of the three moments $\E[1]=1$, $\E[\cos2\vartheta]=C_{11}-C_{22}=:a$ and $\E[\sin2\vartheta]=2C_{12}=:b$. Spectra with elevation need a fourth scalar moment constraint and are not covered by this section. The certificate is therefore an analysis tool for planar spectra. The three-dimensional evaluations of Section~\ref{sec:eval} use the simpler quadratic rule that minimizes $\boldsymbol r^{\mathsf T}C\boldsymbol r$. The certificate does not enter Algorithm~\ref{alg:sel}. It shows when second-order information decides the sign of the mean correlation of a planar spectrum. We do not report how many candidates of the ray-traced data it would certify, and Appendix~\ref{app:synth} reports its coverage on synthetic spectra.

\begin{proposition}[Dual moment certificate]\label{prop:cert}
Let $\boldsymbol y=(y_0,y_1,y_2)$ satisfy $y_0+y_1\cos2\vartheta+y_2\sin2\vartheta\ge\cos(k\boldsymbol u(\vartheta)^{\mathsf T}\boldsymbol r)$ for all $\vartheta$. Then every normalized planar angular measure with second-moment parameters $(a',b')$ and every position error $\|\boldsymbol\xi\|\le\eta$ satisfy
\begin{equation}\label{eq:cert}
c(\boldsymbol r+\boldsymbol\xi)\le y_0+y_1a'+y_2b'+k\eta .
\end{equation}
For $(a',b')$ in the interior of the moment set, the minimum of the right side over feasible $\boldsymbol y$ at $\eta=0$ equals the supremum of $c(\boldsymbol r)$ over all measures with these moments, and the supremum is attained by a measure with at most three atoms, by the duality theory of conic linear problems~\cite{shapiro2001duality}. The lower bound follows by reversing the inequality and subtracting $k\eta$. The dual formulation itself is standard. What we add is its use for the mean cosine at a displaced port, the tolerances in the second moment and in the position, and a numerical certificate that covers a region of second moments.
\end{proposition}
\begin{proof}
Integrating the constraint against the measure gives $c(\boldsymbol r)\le\boldsymbol y^{\mathsf T}(1,a',b')$. The position error changes $c$ by at most $k\|\boldsymbol\xi\|$ because $|\cos x-\cos z|\le|x-z|$ and $|\boldsymbol u^{\mathsf T}\boldsymbol\xi|\le\|\boldsymbol\xi\|$. For an interior moment vector, strong duality holds for this semi-infinite linear program on the compact circle, and the optimal measure lies on a face of the moment cone spanned by at most three of its extreme points, each of which is a point mass.
\end{proof}

The feasibility of $\boldsymbol y$ does not depend on $(a',b')$. One certificate computed at an estimate $\hat C$ therefore covers every $C'$ with $\|C'-\hat C\|_{\rm op}\le\varepsilon$, since this implies $|a'-\hat a|,|b'-\hat b|\le2\varepsilon$:
\begin{equation}\label{eq:robust}
c(\boldsymbol r+\boldsymbol\xi)\le\boldsymbol y^{\mathsf T}(1,\hat a,\hat b)+2\varepsilon(|y_1|+|y_2|)+k\eta .
\end{equation}

\begin{figure}[t]\centering
\includegraphics[width=\columnwidth]{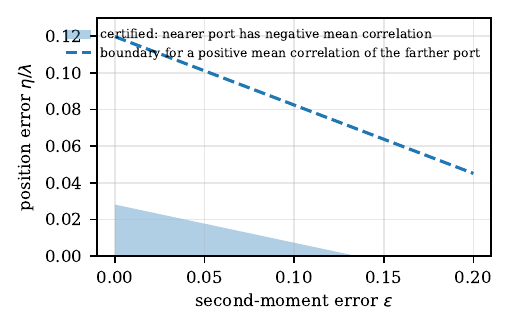}
\caption{Region of second-moment error $\varepsilon$ and position error $\eta$ in which the certificate~\eqref{eq:robust} keeps the mean correlation of the nearer example port negative. The dashed line bounds the region in which the farther port keeps a positive mean correlation.}\label{fig:cert}
\end{figure}

\emph{Example.} Let $C=\mathrm{diag}(0.9,0.1)$, $\boldsymbol r_1=(0.30\lambda,0)$ and $\boldsymbol r_2=(0,0.36\lambda)$. For every planar spectrum with this second moment, the nearer residual $\boldsymbol r_1$ has $c\in[-0.2158,-0.1781]$ and the farther residual $\boldsymbol r_2$ has $c\in[0.7549,0.8363]$. A fourth-order Taylor bound gives the wider intervals $[-0.5989,-0.1254]$ and $[0.7441,0.8533]$. The certificate~\eqref{eq:robust} keeps the mean correlation of $\boldsymbol r_1$ negative for $1.31\varepsilon+2\pi\eta/\lambda<0.1767$, which tolerates $\varepsilon<0.134$ or $\eta<0.028\lambda$ on its own (Fig.~\ref{fig:cert}). The margin $0.1767$ belongs to a dual vector whose feasibility is verified on a grid of 7200 angles with a Lipschitz correction and rechecked on a grid ten times finer, and it is slightly smaller than the exact value $0.1781$. The upper end $-0.1781$ is attained by the spectrum with mass $0.45$ on $\pm\boldsymbol e_1$ and $0.05$ on $\pm\boldsymbol e_2$.

\section{What Information a Safe Choice Needs}\label{sec:stats}
\subsection{Reversal Risk and the Mean Correlation}
A statistics-based rule scores each candidate by a functional of $\mu$. The mean real correlation $c(\boldsymbol r)$ is one such score. The probability of a reversal is a different objective.

\begin{lemma}\label{lem:risk}
Let $(h_f,h_r)$ be zero-mean proper complex Gaussian with $\E|h_f|^2=\E|h_r|^2=1$ and $\E[h_rh_f^*]=\phi$ with $|\phi|<1$. Then
\begin{equation}\label{eq:risk}
\Pr\big(\R(h_rh_f^*)<0\big)=\frac12\Big(1-\frac{\R\phi}{\sqrt{1-(\I\phi)^2}}\Big).
\end{equation}
\end{lemma}
\begin{proof}
Write $h_r=\phi h_f+\sqrt{1-|\phi|^2}\,w$ with $w\sim\mathcal{CN}(0,1)$ independent of $h_f=\rho e^{j\psi}$. Then $\R(h_rh_f^*)=\rho^2\R\phi+\rho\sqrt{1-|\phi|^2}\,\R(we^{-j\psi})$, and $\R(we^{-j\psi})\sim\mathcal N(0,1/2)$ independently of $\rho$. Given $\rho$, a reversal has probability $Q(a\rho)$ with $a=\sqrt2\,\R\phi/\sqrt{1-|\phi|^2}$. Averaging over the Rayleigh density $2\rho e^{-\rho^2}$ gives $\frac12\big(1-a/\sqrt{a^2+2}\big)$, and $a^2+2=2(1-(\I\phi)^2)/(1-|\phi|^2)$ turns this into~\eqref{eq:risk}.
\end{proof}
Under the same model $\E[h_r\mid h_f]=\phi h_f$, and the decomposition in the proof gives $\E|h_r/h_f|\le|\phi|+\sqrt{1-|\phi|^2}\,\pi/2<\infty$ (Appendix~\ref{app:s}), so the mean alignment factor equals the mean real correlation,
\begin{equation}\label{eq:Es}
\E[s]=\R\,\E[h_r/h_f]=\R\phi=c .
\end{equation}
For such channels the certificate of Section~\ref{sec:cert} therefore also bounds the mean of the equalized factor. For non-Gaussian channels, such as two paths with fixed amplitudes and random phases, $\E[s]$ and $c$ can have different signs. For $|\phi|=1$ the ratio $h_r/h_f$ is deterministic, and a reversal occurs with probability one if $\R\phi<0$ and zero otherwise. A candidate with a smaller real correlation can carry a smaller reversal risk when its correlation has a large imaginary part. For example $\phi=0.5$ gives a risk of $0.25$ and $\phi=0.4+0.8j$ gives $1/6$. In our street-level data, choosing by~\eqref{eq:risk} and choosing by $c$ differ by less than one percentage point in reversal rate.

\subsection{The Second Moment and Beyond}
\begin{proposition}[The second moment does not rank ports]\label{prop:insuff}
Let $\mu_1$ put mass $1/4$ on each of the directions $0,\pi/2,\pi,3\pi/2$ and $\mu_2$ the same measure rotated by $\pi/4$. Both have $C=I/2$. For the triangular lattice with $d=1.2\lambda$ and $\boldsymbol D=(0.25\lambda,0)$, the four nearest candidates have $c=(0.500,0.976,0.191,0.191)$ under $\mu_1$ and $c=(0.444,-0.472,-0.002,-0.002)$ under $\mu_2$.
\end{proposition}
The candidate with the largest mean correlation under $\mu_1$ has a negative mean correlation under $\mu_2$, although $C$ is the same. The certificate of Section~\ref{sec:cert} is the safe use of $C$, and ranking ports requires information beyond $C$ in general.

How much is lost depends on the spectra. In synthetic two-cluster von Mises spectra at $d=0.9\lambda$ (Appendix~\ref{app:synth}), choosing the candidate with the largest second-moment lower bound leaves 25.6 percent of the choices with a negative mean correlation against 15.1 percent for the best candidate, and choosing by the midpoint of the bounds that also use the two fourth-order trigonometric moments brings the rate to 15.2 percent. In the five 3GPP CDL profiles the second-moment bound rule is within 0.7 points of the best candidate when the profiles are pooled, and the fourth moments add at most half a point (Section~\ref{sec:eval}). In the ray-traced street the quadratic rule built from $C$ falls significantly short of the full spectrum only at the largest spacing of the vehicular reproduction, and by 3.3 points at $0.9\lambda$ with pedestrian motion (Section~\ref{sec:eval} and Appendix~\ref{app:veh}).

\subsection{What Collinear Observations Cannot Reveal}
A moving terminal observes its forward channel along its own trajectory. Over a short span this trajectory is a straight line, and the history used in Section~\ref{sec:ckm} lies on it. Let the motion direction be $\boldsymbol e_1$ and let $M$ be the reflection $(u_1,u_2)\mapsto(u_1,-u_2)$ across it.

\begin{proposition}[Mirror ambiguity of collinear observations]\label{prop:mirror}
Let the channel be a static, zero-mean, proper complex Gaussian field over the horizontal plane whose covariance is generated by a planar angular measure $\mu$ through~\eqref{eq:phi}, and let it be observed, with independent additive noise of a fixed law, at any finite set of points on the line $\{t\boldsymbol e_1\}$, chosen in advance or adaptively from past observations on that line. Then the observations have the same law under $\mu$ and under its mirror image $M\mu$, whereas
\begin{equation}\label{eq:mirror}
c_{M\mu}(\boldsymbol r)=c_\mu(M\boldsymbol r),\qquad C_{12}(M\mu)=-C_{12}(\mu).
\end{equation}
Consequently, if a prior gives the two mirror images the weights $\pi_0$ and $\pi_1=1-\pi_0$, and a rule must choose between exactly the two candidates $\boldsymbol r$ and $M\boldsymbol r$, whose mean correlations have opposite signs, then every rule that uses only these observations chooses the candidate with the negative mean correlation with probability at least $\min(\pi_0,\pi_1)$.
\end{proposition}
The proof is given in Appendix~\ref{app:mirror}. The conditions matter. Motion of the terminal transverse to the observation line, a channel whose space-time covariance is not mirror invariant, or observations off the line can break the ambiguity, and a third candidate that is safe under both images escapes the bound. Within these conditions the proposition says that the history along the trajectory determines at most the mirror-even part of the spectrum, such as $C_{11}-C_{22}$, and leaves the sign of $C_{12}$ to other information. In the rules of Section~\ref{sec:ckm} this information comes from the surveyed path directions or from the direction of the base station. A rule of the five-city experiment that fits the history only along the motion axis is worse than the nearest port at $0.7\lambda$ and $0.9\lambda$ (Appendix~\ref{app:cities}). That rule also suffers from axial aliasing at a history spacing of $0.56\lambda$, and we did not separate the two causes. A synthetic check of the mechanism is reported in Appendix~\ref{app:mirror}.

\section{Knowledge Maps for Unsurveyed Positions}\label{sec:ckm}
A channel knowledge map stores location-dependent channel information at surveyed positions and interpolates it elsewhere~\cite{zeng2024ckm}. We compare the following ways to score the candidates at a position $\boldsymbol x$ that was not surveyed. All methods receive the same survey, the same base-station position and the same motion direction. Their configurations are listed in Appendix~\ref{app:impl}.
\begin{itemize}
\item \emph{Inverse-distance mixture (IDW).} The angular spectra of the $K$ nearest survey points are mixed with weights proportional to $\mathrm{dist}^{-p}$, and the mixture scores the candidates by $c$ or by~\eqref{eq:risk}. For fixed weights, mixing the spectra is equivalent to mixing the linear scores $c$ and $\phi$. The risk is evaluated after mixing the complex correlations.
\item \emph{Virtual-source transport (VS).} Each surveyed path with arrival direction $\boldsymbol u$ and delay $t$ defines a virtual source at $\boldsymbol x_s+\boldsymbol u\,c_0t$. From the query position the path then arrives from the direction of that source, with its power scaled by the squared ratio of path lengths. This accounts for the parallax that IDW ignores. We do not isolate this effect from that of the history fit.
\item \emph{Virtual-source transport with history fit (VS-Hist).} The terminal has the forward channel at $M$ earlier positions along its trajectory, measured on $N_f$ subcarriers. Taking the $L$ strongest transported paths, we fit their complex amplitudes $\beta_l$ to the wideband history by regularized least squares under the model $H_m(f)=\sum_l\beta_le^{-j2\pi ft_l}e^{jk\boldsymbol u_l^{\mathsf T}\Delta\boldsymbol x_m}$, where $f$ is the subcarrier offset and $\beta_l$ absorbs the carrier phase, and we predict the candidate channels by $\sum_l\beta_le^{jk\boldsymbol u_l^{\mathsf T}\boldsymbol r}$. The fit costs $O(MN_fL^2+L^3)$ operations, about $1.6\times10^5$ complex multiply-accumulates for $M=8$, $N_f=32$ and $L=24$. The transported directions carry the lateral information that the collinear history lacks (Proposition~\ref{prop:mirror}), and the history supplies the phases of the paths. The first and last of the eight observations are 280~ms apart, so the fit assumes a phase reference that is kept over this span and known relative positions. A position error of 1~cm along the arrival direction rotates the phase of a path by about 42 degrees at 3.5~GHz. Section~\ref{sec:budget} measures how much of these errors the fit tolerates.
\item \emph{Learned predictors.} A multilayer perceptron receives the position, the direction to the base station, the motion direction, the candidate residual and the IDW and VS scores, and its history version also receives the delay-domain history and the VS-Hist score. It is trained on the survey to predict the sign of $s$ at each candidate.
\end{itemize}

\section{Proposed Port Selector}\label{sec:selector}
\subsection{A Geometry Rule Without Survey}
Many street channels are dominated by a direct path and its ground reflection. Both arrive from the azimuth of the base station and differ in elevation, so their horizontal projections share the direction $\boldsymbol u_0$ of the base station and have lengths $\sin\theta_1$ and $\sin\theta_2$.

\begin{proposition}[Line-of-sight rule]\label{prop:los}
Let the spectrum consist of two paths with azimuth direction $\boldsymbol u_0$, nonnegative powers $w_1,w_2$ and zenith angles $\theta_1,\theta_2$, and let $x_i=\boldsymbol u_0^{\mathsf T}\boldsymbol r_i$ for the candidates. If $\max_i|x_i|<\lambda/(2\max(\sin\theta_1,\sin\theta_2))$, the candidate with the largest $c(\boldsymbol r_i)$ is the one with the smallest $|x_i|$. If $\sin\theta_1>0$, it is also the candidate chosen by maximizing $\cos(k\sin\theta_1\,x_i)$.
\end{proposition}
\begin{proof}
For two paths with a common azimuth, $c(\boldsymbol r_i)=w_1\cos(k\sin\theta_1x_i)+w_2\cos(k\sin\theta_2x_i)$. Under the stated bound both arguments have absolute value below $\pi$, where the cosine decreases in the absolute value of its argument. Both terms, and their sum, are therefore maximized by the smallest $|x_i|$. If $\sin\theta_1=0$, the direct path is vertical and its term does not distinguish the candidates.
\end{proof}

The rule needs the positions of the terminal and the base station and the heading of the terminal, which a terminal obtains from positioning and inertial sensing. It needs no survey and no channel estimate. The condition of Proposition~\ref{prop:los} holds for every candidate with $|x_i|<\lambda/2$, which includes the nearest candidate whenever $d<\sqrt3\lambda/2$. Farther candidates can violate it, and the rule is then applied without this guarantee. When the direct path and its ground reflection dominate, the direction of the base station also tells the two sides of the motion axis apart whenever it is not parallel to the motion, which is the information that Proposition~\ref{prop:mirror} shows the history cannot supply. An absolute position error of 3~m changes its reversal rate at the street-level site by less than half a point (Section~\ref{sec:budget}).

\subsection{The Selector}
Algorithm~\ref{alg:sel} uses a knowledge map when a survey is available. Without a survey it uses a heuristic based on the port spacing. The line-of-sight rule was better than the nearest port at the tested spacings of $0.9\lambda$ and $1.2\lambda$. At $0.5\lambda$ and $0.7\lambda$ it was worse at the Chicago street-level site (Table~\ref{tab:real4}). At $0.7\lambda$ it was also worse when all five-city queries were pooled, although it was better at eight of the ten sites. We therefore keep the nearest port at the two smaller spacings. The five-city experiment did not include $0.5\lambda$, and other spacings are not tested. Which map rule to use, with or without the history fit, is chosen on validation data, since the history fit loses its advantage under diffuse scattering. We use the history fit only when the phase reference and the relative positions of the history are about as accurate as the levels at which its gain was still significant in Section~\ref{sec:budget}. This is a conservative heuristic, not a validated joint threshold.

\begin{algorithm}[t]
\caption{Return-port selection}\label{alg:sel}
\begin{algorithmic}[1]
\Require candidate residuals $\boldsymbol r_1,\dots,\boldsymbol r_4$, port spacing $d$, positions of terminal and base station, heading
\If{a survey near the terminal is available}
  \State compute the map scores of the candidates (IDW or VS, as validated)
  \If{a wideband history with accurate phase and positions is available and the history fit wins on validation data}
    \State transport the surveyed paths and fit their amplitudes to the causal wideband history
    \State choose the candidate with the largest predicted $\R(h_rh_f^*)$
  \Else
    \State choose the candidate with the best validated map score
  \EndIf
\ElsIf{$d\ge0.9\lambda$}
  \State compute the direction of the base station with its elevation
  \State choose the candidate with the largest $\cos(k\boldsymbol u_h^{\mathsf T}\boldsymbol r_i)$
\Else
  \State choose the nearest candidate
\EndIf
\end{algorithmic}
\end{algorithm}

\section{Ray-Traced Evaluation}\label{sec:eval}
\subsection{Setup}
We use the Chicago scene of an open channel-generation pipeline with Sionna RT~2.2~\cite{hoydis2023sionnart} at 3.5~GHz, isotropic single antennas, the shooting-and-bouncing path solver with at most five interactions, specular reflection, diffraction and edge diffraction, and a fixed seed. A fixed seed does not guarantee identical output of the solver across runs, so our conclusions rest on rates over thousands of queries and on the displacement check below. The scene has three base stations, a street-level site 31~m high and rooftop sites 97 and 108~m high. The pedestrian experiments use the street-level site and the rooftop site at 108~m, and the vehicular sweep of Appendix~\ref{app:veh} uses all three. The receivers are 1~m above the ground in Chicago and 1.5~m in the other cities. We read the path coefficients and the absolute delays of the solver, whose coefficients carry no carrier phase, so the narrowband channel is $h=\sum_l a_le^{-j2\pi f_ct_l}$. All channels at the forward position, at the candidate ports and along the history are traced at their exact positions. The materials of the scene have zero scattering coefficient, and we report a sensitivity case with scattering coefficient $0.3$.

Each query is a position on a trajectory of the scene or a lateral offset of $\pm0.75$~m from it. The motion direction follows the trajectory, and the speed is set to $1.2$~m/s, so that $\|\boldsymbol D\|=0.56\lambda$ for $\tau=40$~ms. The lattice is rotated uniformly at random for each query. The reversal rate is the fraction of queries in which the chosen port has $s<0$. \emph{Spec} chooses the candidate with the largest mean real correlation $c$ computed from the paths at the query position, and \emph{Orc} chooses the candidate with the largest realized $s$. Ties go to the candidate with the lower index. Both use information that a terminal does not have and serve as references.

\subsection{Checks on the Ray-Traced Channels}
We checked the simulated channels in three ways. Displacing a receiver by $0.01\lambda$, $0.05\lambda$ and $0.25\lambda$ in a random horizontal direction gives a median of $s$ of 0.999, 0.992 and 0.81 over 50 positions. In the pedestrian experiments all receivers of a query are traced in the same solver call. Adding diffuse scattering changes the reversal rates of the nearest port and of the best candidate by less than one percentage point. The trajectories of the scene carry speeds of about 9~m/s, closer to vehicles than to pedestrians. We therefore also report the vehicular displacement of the trajectories with an unrestricted lattice (Appendix~\ref{app:veh}).

\subsection{Reversal Rates at Street-Level and Rooftop Sites}
Table~\ref{tab:real4} reports 1000 queries per base station. For the street-level site the nearest port reverses the gradient in 12, 20 and 49 percent of the exchanges for spacings of $0.5\lambda$, $0.7\lambda$ and $0.9\lambda$, while at least one candidate is aligned in all but 0.2, 1.6 and 9.2 percent. For the Chicago rooftop site the nearest port reverses in 0.4, 0.2 and 5.8 percent, and the line-of-sight rule with elevation reverses in 0.5 percent at $0.9\lambda$. At the street-level site the line-of-sight rule is worse than the nearest port at $0.5\lambda$ and $0.7\lambda$ and better at $0.9\lambda$. The rooftop site is not safe at every spacing. In the query set of Section~\ref{sec:ckmeval}, at $1.2\lambda$, the nearest port reverses 37.3 percent of its returns there and the line-of-sight rule 9.5 percent.

\begin{table}[t]\centering
\caption{Reversal rate with pedestrian motion and a finite aperture in Chicago (1000 queries per site). The quadratic rule minimizes $\boldsymbol r^{\mathsf T}C\boldsymbol r$ with the horizontal second moment $C$ of the paths at the query position.}\label{tab:real4}
\setlength{\tabcolsep}{4.5pt}
\begin{tabular}{lcccccc}\toprule
 & \multicolumn{3}{c}{Street level (31 m)} & \multicolumn{3}{c}{Rooftop (108 m)}\\
$d/\lambda$ & 0.5 & 0.7 & 0.9 & 0.5 & 0.7 & 0.9\\\midrule
Nearest & 0.122 & 0.196 & 0.485 & 0.004 & 0.002 & 0.058\\
LoS with elevation & 0.165 & 0.257 & 0.347 & 0.002 & 0.003 & 0.005\\
Quadratic rule & 0.102 & 0.159 & 0.273 & 0.002 & 0.001 & 0.003\\
Spec (reference) & 0.100 & 0.148 & 0.240 & 0.002 & 0.001 & 0.004\\
Orc (reference) & 0.002 & 0.016 & 0.092 & 0 & 0 & 0.001\\\bottomrule
\end{tabular}
\end{table}

\begin{figure}[t]\centering
\includegraphics[width=\columnwidth]{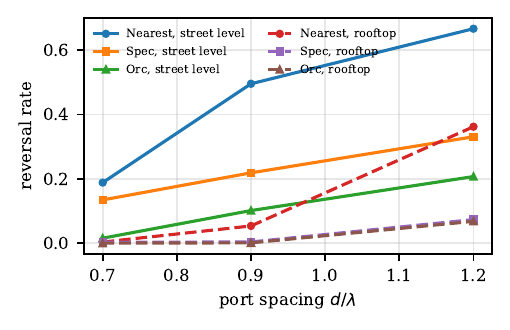}
\caption{Reversal rate of the nearest port, the full-spectrum reference (Spec) and the best of four candidates (Orc) against the port spacing. Pedestrian motion, finite aperture, 9600 queries per site.}\label{fig:spacing}
\end{figure}

\begin{figure}[t]\centering
\includegraphics[width=\columnwidth]{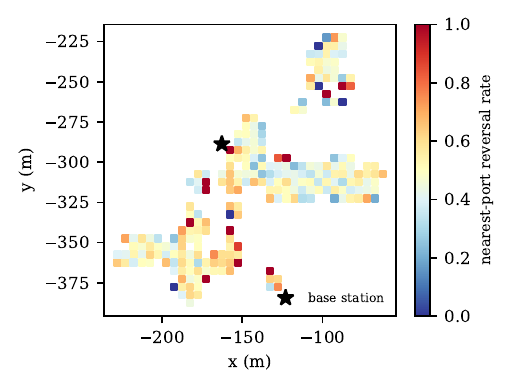}
\caption{Reversal rate of the nearest port at $d=0.9\lambda$ for the street-level site, averaged over $5\times5$~m cells.}\label{fig:map}
\end{figure}

Fig.~\ref{fig:spacing} shows the same ordering of rules for a separate set of 9600 queries per site at spacings of $0.7\lambda$, $0.9\lambda$ and $1.2\lambda$, and Fig.~\ref{fig:map} shows that reversals occur across the streets of the street-level site. At $d=0.5\lambda$ the residual of the nearest port has a median of $0.15\lambda$, well inside the quarter-wavelength radius that certifies a positive mean correlation for every spectrum, and the nearest port still reverses 12 percent of the returns. The full spectrum removes only 2 of these 12 percentage points, whereas at $0.9\lambda$ it removes half of the reversals.


\subsection{Sufficiency of Second-Order Statistics}
For the five 3GPP CDL profiles~\cite{tr38901}, reduced to the horizontal plane and rotated at random, we computed the fraction of geometries in which the chosen port has a negative mean correlation. At $d=0.9\lambda$ this fraction is 38.1 percent for the nearest port, 15.0 percent for the bound rule, which maximizes the midpoint of the second-moment bounds of Proposition~\ref{prop:cert}, and 14.3 percent for the best candidate, averaged over the profiles. The bound rule closes about 97 percent of the gap in these profiles, and adding the fourth moments lowers it to 14.5 percent (Appendix~\ref{app:synth}). It is scored by the fraction of geometries with a negative mean correlation, whereas the quadratic rule below is scored by the rate of single reversals. In the ray-traced street, with vehicular displacements from the trajectories and an unrestricted lattice, the quadratic rule has a reversal rate 7.8 percentage points above the full-spectrum rule at $d=1.2\lambda$ (paired bootstrap interval $[4.6,11.2]$ over 500 queries, without accounting for spatial clustering), while the difference at $0.9\lambda$ is not significant. With pedestrian motion the quadratic rule is 3.3 points above Spec at $0.9\lambda$ (Table~\ref{tab:real4}).

\subsection{Positions Outside the Survey}\label{sec:ckmeval}
For Table~\ref{tab:real3} the trajectories of the street-level site are grouped by the 10~m cell of their first step. The trajectories of 22 cells with 2400 queries form the test set, those of further cells the validation set with 1186 queries, and the remaining 4517 queries, at least 2~m from any test query, the survey. The strongest knowledge-map rule is selected on the validation set before the test set is used. The full results for five survey densities, three spacings, the scattering case and the rooftop site are in Appendix~\ref{app:real3}.

\begin{table}[t]\centering
\caption{Reversal rate at unsurveyed positions, Chicago street-level site, $d=0.9\lambda$.}\label{tab:real3}
\begin{tabular}{lcc}\toprule
Rule & Full survey & 10 m survey\\\midrule
Nearest & 0.494 & 0.494\\
LoS with elevation & 0.366 & 0.366\\
Selected map without history & 0.319 & 0.330\\
Learned without history & 0.318 & 0.318\\
VS-Hist & 0.270 & 0.265\\
Learned with history & 0.267 & 0.259\\
Spec (reference) & 0.215 & 0.215\\
Orc (reference) & 0.090 & 0.090\\\bottomrule
\end{tabular}
\end{table}

The causal history lowers the reversal rate from 0.32 to 0.27 through the physical fit of VS-Hist (paired cluster-bootstrap interval of the difference $[-0.3,11.2]$ percentage points over the 22 test cells, so this gain alone is not significant). The selected maps without history remain 10 to 12 percentage points above Spec, and the history-based rules reduce this gap to 4 to 6 points. At the two primary survey densities, the advantage of the learned predictor with history over VS-Hist has paired 95 percent bootstrap intervals, clustered by 10~m cells of the trajectory starts, of $[-1.2,2.2]$ and $[-0.8,2.4]$ percentage points, below the value threshold of 3 points, which was fixed before the test runs. At the secondary 5~m density, with 79 survey points, the learned predictors without and with history are better than the selected map by 2.7 and 7.4 points (95 percent intervals $[0.9,5.3]$ and $[3.7,12.0]$). At this density the validation set selected the IDW map also for the history option, and VS-Hist itself reaches 0.254. At the 2~m density, with 236 survey points, the learned predictors and the selected rules are within half a point of each other. On the rooftop site with the full survey the learned predictors are worse than the selected map by 0.9 and 2.2 points, and with the 5~m survey of 63 points they fail, with reversal rates of 0.298 and 0.376 against 0.009 for the selected map. With diffuse scattering of coefficient 0.3 the median number of paths grows from 38 to 727, the nearest-port and oracle rates change by less than one percentage point, and VS-Hist reaches 0.379 while the selected map without history reaches 0.328. The choice among map rules uses validation labels from the same ray-traced scene, which a deployment would have to obtain by measurement. All intervals use 2000 resamples of whole clusters with the queries pooled, so larger clusters weigh more, and they describe the sampled sites and cells rather than other cities.

\subsection{Five Cities Without a Survey}
We traced the four city scenes shipped with Sionna RT, namely Munich, Etoile, Florence and San Francisco, with a rooftop and a street-level base station each, at random outdoor positions with the motion along the local street axis. Together with the two Chicago sites this gives 38184 queries. Each city is held out in turn, and learned predictors are trained on the other four cities with a description of the local street geometry given by the free distance in 16 directions. Nothing is surveyed in the held-out city.

\begin{table}[t]\centering
\caption{Zero-survey port selection in five cities, $d=0.9\lambda$, all ten sites pooled. $\E[\bar s]$ is the mean of $s$ clipped to $[-3,3]$.}\label{tab:xscene}
\begin{tabular}{lcc}\toprule
Rule & Reversal rate & $\E[\bar s]$\\\midrule
Nearest & 0.367 & 0.200\\
Street axis & 0.479 & 0.022\\
LoS with elevation & 0.205 & 0.496\\
Learned, other cities & 0.214 & 0.461\\
Learned with history, other cities & 0.225 & 0.423\\
Spec (reference) & 0.144 & 0.603\\
Orc (reference) & 0.068 & 0.827\\\bottomrule
\end{tabular}
\end{table}

\begin{figure*}[t]\centering
\includegraphics[width=\textwidth]{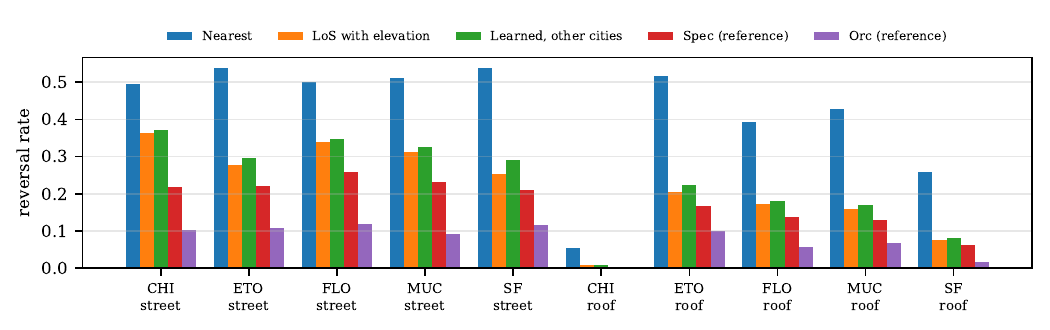}
\caption{Zero-survey reversal rate per site at $d=0.9\lambda$. Each city is held out in turn and the learned predictor is trained on the other four cities.}\label{fig:persite}
\end{figure*}
At $0.9\lambda$ the nearest port reverses between 26 percent (San Francisco rooftop) and 54 percent (Etoile and San Francisco street level) of the returns in the four Sionna cities, and 5.4 percent at the Chicago rooftop site (5.8 percent in the separate query set of Table~\ref{tab:real4}). The line-of-sight rule with elevation has the lowest reversal rate among the deployable rules at every one of the ten sites (Fig.~\ref{fig:persite}). Its advantage over the learned predictor without history is 0.9 percentage points (95 percent interval $[0.7,1.2]$, clustered by 20~m cells, which are nested in sites and cities), and the history version is worse by 2.0 points. At $0.7\lambda$ the learned predictor without history is better by 1.1 points (interval $[0.4,2.0]$), below the threshold of 3 points fixed before the test runs, and at $1.2\lambda$ it is worse by 2.0 points (Appendix~\ref{app:cities}). The tested street-axis rule is worse than the nearest port, whereas the base-station geometry rule is better.

\subsection{Error Budget of the History Fit}\label{sec:budget}
\begin{figure*}[t]\centering
\includegraphics[width=\textwidth]{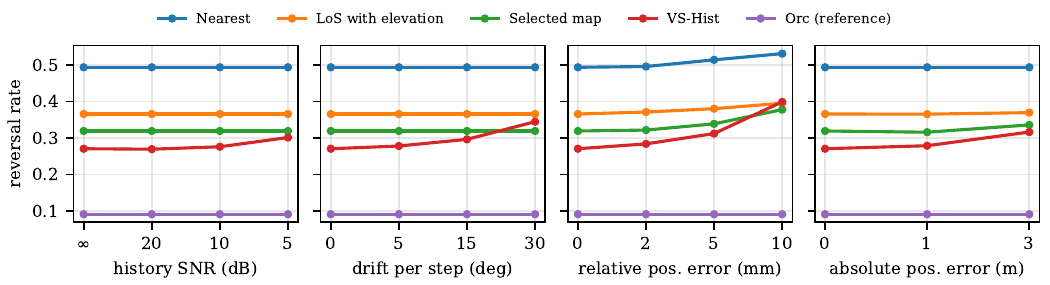}
\caption{Reversal rate at unsurveyed positions under errors in the history and the positions, Chicago street level, $d=0.9\lambda$, full survey. Each panel adds one error, and the other errors are zero.}\label{fig:budget}
\end{figure*}

The results above assume a noise-free history, exact positions and calibrated ports. We repeated the test of Table~\ref{tab:real3} at $0.9\lambda$ with the full survey and with each of six errors added in turn, using the code and the parameters selected for Table~\ref{tab:real3} without retuning. Every query draws its errors once, and all rules see the same draws. The errors are complex Gaussian noise on the history at a given signal-to-noise ratio, a phase random walk across the 40~ms steps of the history, Gaussian errors on the relative positions of the history and on the common displacement of the candidates, a Gaussian error on the absolute position, a rotation of the body offsets by an orientation error, and a fixed phase error of each return port relative to the forward port. The last one changes the true $s$, and all rules, including Orc, are scored against it. Two joint settings combine the errors at moderate and severe levels. Without errors the run reproduces the rates of Table~\ref{tab:real3} exactly. Fig.~\ref{fig:budget} shows four representative errors, and Appendix~\ref{app:budget} lists all settings with paired intervals.

Three statements follow. First, the gain of VS-Hist over the nearest port survives every tested error. It is 22.4 points without errors (95 percent interval $[15.0,28.2]$) and still 12.7 points $[7.8,16.6]$ in the severe joint setting. Second, its gain over the line-of-sight rule, 9.5 points $[1.3,20.9]$ without errors, stays significant for a history signal-to-noise ratio of 10~dB, a phase drift with a standard deviation of $5^\circ$ per step, relative position errors with a standard deviation of 2~mm per axis, an absolute position error of 1~m, an orientation error of $10^\circ$ and a port calibration error of $30^\circ$, each on its own. At the next tested levels of 5~dB, $15^\circ$ per step, 5~mm and 3~m, and in both joint settings, the interval includes zero, although the point estimate favors VS-Hist in every setting except relative errors of 1~cm. Third, its gain over the selected map without history is not significant even without errors, and its point estimate turns negative at a drift of $30^\circ$ per step, at 1~cm and in the severe joint setting. At their largest tested levels, the relative position error and the phase drift degrade VS-Hist the most, in line with the 42 degrees of phase that 1~cm produces. Calibration and orientation errors generally increase the non-oracle reversal rates, although not for every rule at every tested level, and change the differences between rules less. An absolute position error of 3~m weakens the survey-based rules, with the selected map at 0.336 and VS-Hist at 0.316, but barely moves the line-of-sight rule, at 0.370. The gain over the line-of-sight rule therefore remained significant at the tested single-error levels of $5^\circ$ per 40~ms step and 2~mm per axis, both expressed as standard deviations, but not at the next tested levels or in either joint setting. We did not test intermediate levels or estimate a critical value, and the loss of significance does not mean that the line-of-sight rule is better. In every tested setting the nearest port was the worst deployable rule at this spacing.

\subsection{Effect on Training}\label{sec:train}
In a surrogate in which the alignment factor is drawn independently of the minibatch gradient and the gradient noise is unbiased, the drift of stochastic gradient descent along the exact gradient scales with the mean of the factor~\cite{bottou2018sgd}. In general the factor and the gradient are dependent, and this scaling need not hold. Table~\ref{tab:align} reports the mean of $s$ clipped to $[-3,3]$ for the rules of Table~\ref{tab:real3}. At $0.9\lambda$ the nearest port leaves a mean clipped alignment of 0.03, and at $1.2\lambda$ it is negative. VS-Hist raises it to 0.42 and 0.31. A time to a target accuracy that grows as the inverse of the alignment, as observed for a related setup~\cite{li2026keep}, would translate these values into large differences in training speed. We do not test this prediction here.

\begin{table}[t]\centering
\caption{Mean of the clipped alignment $\bar s=\min(\max(s,-3),3)$ at unsurveyed positions (Chicago street level, full survey).}\label{tab:align}
\begin{tabular}{lccc}\toprule
$d/\lambda$ & 0.7 & 0.9 & 1.2\\\midrule
Nearest & 0.496 & 0.031 & $-0.226$\\
LoS with elevation & 0.390 & 0.240 & 0.219\\
Selected map without history & 0.515 & 0.325 & 0.273\\
VS-Hist & 0.569 & 0.417 & 0.306\\
Learned with history & 0.563 & 0.408 & 0.299\\
Spec (reference) & 0.618 & 0.501 & 0.349\\
Orc (reference) & 1.020 & 0.868 & 0.721\\\bottomrule
\end{tabular}
\end{table}

\begin{figure}[t]\centering
\includegraphics[width=\columnwidth]{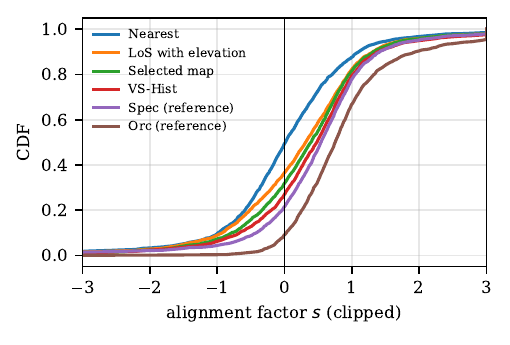}
\caption{Distribution of the clipped alignment factor of the chosen port at unsurveyed positions, Chicago street level, $d=0.9\lambda$, full survey.}\label{fig:cdf}
\end{figure}
Fig.~\ref{fig:cdf} shows that the rules differ in the distribution of the clipped alignment factor beyond its sign. We trained classifiers in two settings, both with the paired test queries of Table~\ref{tab:real3} at $0.9\lambda$ and with the learning rate selected on a validation split for each rule.

The first setting is a surrogate. A linear classifier on the digits data set is trained for 3000 steps, and each update multiplies the exact minibatch gradient by a value of $\bar s$ drawn from the test queries. Every rule other than the nearest port reached a mean test accuracy between 94.7 and 95.2 percent, close to the 95.1 percent without aging. The nearest port reached 87.8 percent on average and between 81.2 and 92.9 percent across five seeds. The surrogate has only an additive gradient noise, and the clipping of the factor belongs to the surrogate, not to the return.

The second setting runs the scalar link of Section~\ref{sec:model} for every sample. The forward link has a signal-to-noise ratio of 20~dB and a channel estimate with a normalized error of $-20$~dB, the terminal equalizes with the estimate and returns $\delta^*$ by truncated channel inversion, skipping samples whose estimated channel power is below 0.1, and the return link adds noise at 20 or 10~dB relative to the mean power of the returned symbols. The forward and return channels are the paired ray-traced channels of the forward port and of the port each rule chooses. The exact gradient is taken along the actual forward computation with the estimated channel, holding the realized noise and channel estimate fixed, so it carries the factor $\R(h_f/\hat h_f)$ for each sample. Table~\ref{tab:d1} reports the result. The nearest port is about 13 to 15 points below every other rule at both return signal-to-noise ratios, and its accuracy varies far more across seeds. Excluding the nearest port, the mean accuracies span 1.5 points at 20~dB and 1.7 points at 10~dB, including Orc and the ideal return. The intervals of the line-of-sight rule, the selected map and VS-Hist overlap, and we did not compute paired intervals of their differences. The cosine between the returned and the exact gradient separates them. It is ordered exactly opposite to the reversal rate, with a Spearman coefficient of $-1$ at both signal-to-noise ratios. The projection of the returned gradient on the exact one, divided by the squared norm of the exact gradient, is the step taken along the exact gradient per unit learning rate. The mean across seeds of its per-run median is 0.385 for VS-Hist against 0.225 for the line-of-sight rule at 20~dB. On this small convex task the final accuracies give no evidence of a difference among the deployable rules other than the nearest port. Whether the larger step along the exact gradient becomes faster convergence on a harder task with a fixed budget of steps is a hypothesis we did not test.

\begin{table}[t]\centering
\caption{Training through the scalar link with truncated channel inversion (binary digits task, linear classifier, 2000 steps). Test accuracy in percent, mean and 95 percent interval over ten seeds, and at a return signal-to-noise ratio of 20~dB the mean cosine between the returned and the exact gradient and the median projection coefficient.}\label{tab:d1}
\setlength{\tabcolsep}{2.6pt}
\begin{tabular}{lcccc}\toprule
Rule & Return 20 dB & Return 10 dB & Cosine & Proj.\\\midrule
Nearest & 76.8 [72.2, 81.3] & 76.2 [71.6, 80.9] & 0.037 & 0.023\\
LoS with elevation & 89.5 [88.7, 90.3] & 89.5 [88.7, 90.4] & 0.260 & 0.225\\
Selected map & 90.3 [89.4, 91.1] & 90.4 [89.5, 91.2] & 0.357 & 0.296\\
VS-Hist & 90.5 [90.1, 90.9] & 90.8 [90.3, 91.2] & 0.446 & 0.385\\
Orc (reference) & 91.0 [90.4, 91.6] & 91.2 [90.6, 91.8] & 0.734 & 0.688\\
Ideal, $h_r=h_f$ & 90.6 [89.8, 91.4] & 91.1 [90.5, 91.7] & 0.941 & 0.978\\\bottomrule
\end{tabular}
\end{table}

\section{Conclusion}
At a port spacing of $0.9\lambda$ in ray-traced streets, the port nearest to the forward position reverses a large fraction of the returned gradients, at street-level sites in every city we traced and at rooftop sites in four of them. A certificate from second moments decides when the mean real correlation of a port is safely positive or negative, location statistics indicate which port is likely to be aligned, and the causal history of the forward channel moves the choice toward the realization. Under a static Gaussian model, the history along a straight trajectory cannot tell the two sides of the motion axis apart, so the side must come from other information, such as a map or the direction of the base station. Without a survey and at $0.9\lambda$, the direction of the base station with its elevation had the lowest reversal rate among the deployable rules at all ten sites. In the primary settings the multilayer perceptrons we tested did not improve on the physical rules by the margin fixed before the test runs, while with a sparse survey of 79 points the learned predictor with history was better than the selected map by 7.4 points, and on the rooftop site with a sparse survey the learned predictors failed. The gain of the history fit over the line-of-sight rule remained significant at the tested single-error levels of $5^\circ$ of phase drift per step and 2~mm of position error but not at the next tested levels, while its gain over the nearest port survived every tested error. On a power-limited scalar link the nearest port costs about 13 to 15 points of test accuracy, and the gradient cosine of the rules follows their reversal rates. A gap to the realization-aware oracle remains, and its causes are not isolated here. Measured channels, rotation of the terminal during the exchange, directional antennas, coupling between ports, convergence speed on harder tasks, and probing that would reach the realization are left for future work.

\appendices
\section{The Alignment Factor}\label{app:s}
With $y=h_fq+n$ and the equalized output $z=\R(y/h_f)$, the loss $\mathcal L(z)$ is real and $\partial z/\partial y^*=1/(2h_f^*)$, so $\delta=2\,\partial\mathcal L/\partial y^*=\mathcal L'(z)/h_f^*$. The derivative of the loss with respect to the real symbol is $\partial\mathcal L/\partial q=\R(h_f^*\delta)=\mathcal L'(z)$. The terminal returns $\delta^*$ through the reciprocal channel and the transmitter conjugates the received $h_r\delta^*$, which gives $h_r^*\delta$ and $\R(h_r^*\delta)=\mathcal L'(z)\,\R(h_r^*/h_f^*)=\mathcal L'(z)\,s$, because $\R(h_r^*/h_f^*)=\R(h_rh_f^*)/|h_f|^2$. Sending $\delta$ directly through the reciprocal channel would instead give $\R(h_r\delta)$, which need not equal the gradient analyzed here. Multiplying by $\boldsymbol x$ shows that the returned gradient is $s$ times the exact gradient plus the backward noise. The derivation requires a real symbol, an equalizing receiver with a nonzero forward channel and a real loss of the equalized output. Without equalization the ratio of the returned to the exact derivative is $\R(h_r^*\delta)/\R(h_f^*\delta)$, which depends on the phase of $\delta$.

\emph{Integrability under the Gaussian model.} With the decomposition of the proof of Lemma~\ref{lem:risk}, $h_r/h_f=\phi+\sqrt{1-|\phi|^2}\,w/h_f$, where $w$ and $h_f$ are independent. The modulus $|w|$ is Rayleigh distributed with $\E|w|=\sqrt\pi/2$, and $\E[1/|h_f|]=\int_0^\infty2e^{-\rho^2}d\rho=\sqrt\pi$. Hence $\E|w/h_f|=\pi/2$ and $\E|h_r/h_f|\le|\phi|+\sqrt{1-|\phi|^2}\,\pi/2$. Since $w$ has zero mean and is independent of $h_f$, $\E[w/h_f]=0$ and $\E[h_r/h_f]=\phi$, which gives~\eqref{eq:Es}. The second moment of $w/h_f$ involves $\E[1/|h_f|^2]=\int_0^\infty2\rho^{-1}e^{-\rho^2}d\rho$, which diverges, so $s$ has infinite variance whenever $|\phi|<1$.

\section{Synthetic Spectra and the 3GPP CDL Profiles}\label{app:synth}
\emph{Two-cluster spectra.} Each synthetic planar spectrum mixes two von Mises clusters. The main cluster has a uniform direction $\vartheta_1$, the second cluster the direction $\vartheta_1+\pi/2$ plus a Gaussian perturbation of standard deviation 0.3, the main cluster carries a power fraction drawn from $U(0.6,0.95)$, and each concentration is drawn from $U(5,50)$. The spectra are asymmetric, so $\phi$ is complex. The candidates are the four lattice points nearest to a displacement of uniform direction and radius drawn from $U(0.2,0.6)\,(d/0.5\lambda)\,\lambda$. An estimated second moment $\hat C=C+E$ is formed with a random traceless symmetric $E$ of operator norm $\varepsilon$, projected back to the feasible set. The rules are the nearest candidate (Euclidean), the quadratic rule with $\hat C$, the certificate rule that maximizes the robust lower bound~\eqref{eq:robust} at $\hat C$, and the oracle that maximizes the true $c$. A choice counts as wrong when the true $c$ of the chosen candidate is negative. All integrals use a grid of 1440 angles.

\begin{table}[t]\centering
\caption{Two-cluster synthetic spectra, 2000 scenes per row. Fraction of choices with negative mean correlation, and the number of scenes in which the certificate rule chose a candidate with a positive lower bound. No such candidate had a negative true $c$.}\label{tab:synth}
\setlength{\tabcolsep}{3.5pt}
\begin{tabular}{ccccccc}\toprule
$d/\lambda$ & $\varepsilon$ & Euclid & Quadratic & Certificate & Oracle & Certified\\\midrule
0.5 & all & 0 & 0 & 0 & 0 & 2000\\
0.7 & 0 & 0.148 & 0.047 & 0.048 & 0.030 & 1808\\
0.7 & 0.05 & 0.148 & 0.050 & 0.057 & 0.030 & 1645\\
0.7 & 0.1 & 0.148 & 0.057 & 0.081 & 0.030 & 1489\\
0.9 & 0 & 0.340 & 0.226 & 0.222 & 0.144 & 1284\\
0.9 & 0.05 & 0.340 & 0.232 & 0.245 & 0.144 & 1090\\
0.9 & 0.1 & 0.340 & 0.237 & 0.272 & 0.144 & 945\\\bottomrule
\end{tabular}
\end{table}

Table~\ref{tab:synth} shows three things. In this synthetic sample port selection is no problem at $0.5\lambda$. The certificate never certified a candidate whose true mean correlation was negative, which is what Proposition~\ref{prop:cert} guarantees, and the count of certified scenes shrinks as $\varepsilon$ grows. At $0.9\lambda$ both second-moment rules stop near 0.22 while the oracle reaches 0.14. The tested second-moment rules thus leave a gap for these spectra, and a larger $\varepsilon$ makes the certificate rule more conservative than the quadratic rule.

\emph{Fourth moments.} In a fresh set of 2000 scenes at $d=0.9\lambda$ and $\varepsilon=0$, we added the exact moments $\E[\cos4\vartheta]$ and $\E[\sin4\vartheta]$ to the constraints of the linear program, solved on the same angle grid. Choosing by the lower bound from the second moment alone leaves 25.6 percent of the choices wrong. Choosing by the lower bound with the fourth moments leaves 16.2 percent, choosing by the midpoint of the fourth-moment bounds 15.2 percent, and the oracle 15.1 percent, so the fourth moments recover about 99 percent of the gap. The fourth-moment bounds are exact for spectra on the grid and carry no Lipschitz correction for continuous spectra, unlike the second-moment certificate, and the scene set differs from that of Table~\ref{tab:synth}, so the two are compared only within each set.

\emph{Mirror parity of the moments.} The identities $\cos4\vartheta=8u_1^4-8u_1^2+1$ and $\sin4\vartheta=4u_1u_2(u_1^2-u_2^2)$ show that $\cos4\vartheta$ depends only on $u_1$ and is mirror even, whereas $\sin4\vartheta$, like $\sin2\vartheta=2u_1u_2$, changes sign under the mirror of Proposition~\ref{prop:mirror}. In 1000 scenes at $0.9\lambda$, choosing by the midpoint of the bounds gives 0.183 wrong choices with the second moment, 0.167 when $\E[\cos4\vartheta]$ is added and 0.157 when $\E[\sin4\vartheta]$ is added as well, equal to the oracle. The second-moment rule already contains the mirror-odd $\E[\sin2\vartheta]$, so this sequence shows the increments of the fourth moments within this family of spectra and is not an information-theoretic ceiling of second-order rules.

\begin{table}[t]\centering
\caption{3GPP CDL profiles reduced to the plane and rotated at random, 1000 geometries per entry. Fraction of choices with negative mean correlation. M2 is the bound rule from the second moment, M2c4 adds $\E[\cos4\vartheta]$ and M4 adds both fourth-order moments.}\label{tab:cdl}
\setlength{\tabcolsep}{4pt}
\begin{tabular}{llccccc}\toprule
$d/\lambda$ & Profile & Nearest & M2 & M2c4 & M4 & Best\\\midrule
0.5 & A, B, C & 0 & 0 & 0 & 0 & 0\\
0.5 & D, E & 0.009, 0.010 & 0 & 0 & 0 & 0\\
0.7 & A & 0.141 & 0.027 & 0.026 & 0.026 & 0.026\\
0.7 & B & 0.042 & 0.025 & 0.025 & 0.025 & 0.025\\
0.7 & C & 0.050 & 0.025 & 0.025 & 0.025 & 0.025\\
0.7 & D & 0.236 & 0.043 & 0.043 & 0.043 & 0.043\\
0.7 & E & 0.235 & 0.029 & 0.029 & 0.029 & 0.029\\
0.9 & A & 0.349 & 0.163 & 0.159 & 0.153 & 0.152\\
0.9 & B & 0.398 & 0.157 & 0.152 & 0.151 & 0.149\\
0.9 & C & 0.386 & 0.169 & 0.159 & 0.159 & 0.153\\
0.9 & D & 0.392 & 0.131 & 0.131 & 0.131 & 0.131\\
0.9 & E & 0.381 & 0.129 & 0.129 & 0.129 & 0.129\\
0.9 & pooled & 0.381 & 0.150 & 0.146 & 0.145 & 0.143\\\bottomrule
\end{tabular}
\end{table}

\emph{CDL profiles.} Table~\ref{tab:cdl} lists the CDL results of Section~\ref{sec:eval} per profile. In the line-of-sight profiles D and E the nearest port is already wrong in about 24 percent of the geometries at $0.7\lambda$, against 3 to 4 percent for the second-moment rule. The moments of the CDL spectra are computed on an angle grid with periodic linear interpolation, so these bounds are numerical approximations. In these standard profiles the second moment is nearly sufficient for the tested rule, and the fourth-moment gain seen in the two-cluster spectra does not carry over. The reduction neglects the zenith angles, the coupling between ports, and the orientation of the terminal, which the random rotation replaces by a uniform orientation.

\section{The Mirror Ambiguity}\label{app:mirror}
\emph{Proof of Proposition~\ref{prop:mirror}.} At the points $t_i\boldsymbol e_1$ the covariance of the field is $K_{ij}=\int e^{jku_1(t_i-t_j)}d\mu$, which depends on $\mu$ only through the law of $u_1$. The pseudo-covariance vanishes and the mean is zero, so this covariance fixes the joint law of the field at the points. The reflection $M$ keeps $u_1$, so $\mu$ and $M\mu$ give the same law of $u_1$ and therefore the same law of the field on the line, with or without the independent noise. An adaptive rule that picks the next point on the line from the past observations induces the same law of the whole record under both measures, by induction over the points. The first identity of~\eqref{eq:mirror} follows from $\boldsymbol u^{\mathsf T}M\boldsymbol r=(M\boldsymbol u)^{\mathsf T}\boldsymbol r$, and the second from $C_{12}=\int u_1u_2\,d\mu$. If $c_\mu(\boldsymbol r)$ and $c_\mu(M\boldsymbol r)$ have opposite signs, the safe candidate under $\mu$ is the unsafe one under $M\mu$. Since the record has the same law under both, any rule, randomized or not, chooses $\boldsymbol r$ with the same probability $q$ under both, and its probability of an unsafe choice is $\pi_0(1-q)+\pi_1q\ge\min(\pi_0,\pi_1)$ when $\boldsymbol r$ is the safe candidate under $\mu$.

\emph{Conditions.} The bound is a population risk for the choice between the two candidates under the stated prior. With a context variable $Z$ that is independent of the record given the hypothesis, it becomes $\E_Z\min(\Pr(\mu\mid Z),\Pr(M\mu\mid Z))$. A field $H(\boldsymbol x,t)=Ze^{jk\boldsymbol u^{\mathsf T}(\boldsymbol x+\boldsymbol vt)}$ observed at one point over time, with $\boldsymbol u=(0.6,0.8)$ and a transverse velocity $\boldsymbol v=(0,0.25\lambda)$ per time unit, has lag-one covariances $0.309\pm0.951j$ under the two mirror images, so transverse motion breaks the ambiguity. Observations off the line can break it. Two points, one on each of two perpendicular axes, already give a cross covariance at the oblique lag $(h,-h)$. For $\mu=\delta_{(0.6,0.8)}$ and $h=0.1\lambda$ its real part is 0.992 under $\mu$ and 0.637 under $M\mu$, and the expansion $c(h,-h)=1-\frac{k^2h^2}{2}(C_{11}+C_{22}-2C_{12})+o(h^2)$ shows that such a lag carries $C_{12}$. The proposition also does not claim that the magnitude of $C_{12}$ is unconstrained. For some families of spectra it follows from the law of $u_1$.

\emph{A synthetic check.} We tested the mechanism with the spectrum $\mu_\theta=0.8\cdot\frac12(\delta_{\boldsymbol u(\theta)}+\delta_{-\boldsymbol u(\theta)})+0.2\cdot$uniform, with $|\theta|$ uniform on $(0.15,\pi/2-0.15)$ and a sign of $\theta$ that is $\pm1$ with equal probability. Two ports at $0$ and $0.2\lambda$ along $\boldsymbol e_1$ observe eight independent snapshots with noise variance 0.1. Their covariance depends only on $\cos\theta$ and is exactly mirror invariant. A context bit equals the sign of $\theta$ with probability 0.9. The two candidates are $\boldsymbol r=(0.25,0.25)\lambda$ and $M\boldsymbol r=(0.25,-0.25)\lambda$, and in all 5000 test scenes they have mean correlations of opposite signs. A Bayes rule with a mirror-symmetric prior that ignores the context chooses the unsafe candidate in 0.494 of the scenes, the Bayes rule with the true context prior in 0.092, a multilayer perceptron trained on 20000 scenes with the sample covariance and the context in 0.092, and the same perceptron with the context permuted at test time in 0.499. The context here is an artificial bit that a parametric Bayes rule models exactly, so the check confirms the mechanism and the bound, not an advantage of learning.

\section{Vehicular Displacements in Chicago}\label{app:veh}
The trajectories of the Chicago scene move at about 9~m/s, so over $\tau=40$~ms the 10th, 50th and 90th percentiles of the displacement are about $2\lambda$, $4\lambda$ and $6\lambda$, far beyond the pedestrian $0.56\lambda$. For these displacements we used a local lattice of $9\times9$ points around $-\boldsymbol D$ in the body frame, so that the four candidates are the nearest ports, without an aperture limit. The geometry rule here uses only the azimuth of the base station, and Spec and the quadratic rule are computed from the paths at the query position.

\begin{table}[t]\centering
\caption{Vehicular displacements in Chicago, unrestricted lattice, 1000 queries per site and spacing. LoS uses the azimuth only. The median numbers of paths are 20, 37 and 21.}\label{tab:veh}
\setlength{\tabcolsep}{3.2pt}
\begin{tabular}{llccccc}\toprule
Site & $d/\lambda$ & Nearest & LoS & Quadratic & Spec & Orc\\\midrule
Rooftop 108 m & 0.9 & 0.034 & 0.030 & 0.005 & 0.004 & 0\\
 & 1.2 & 0.144 & 0.464 & 0.026 & 0.027 & 0.023\\
 & 1.5 & 0.246 & 0.589 & 0.078 & 0.078 & 0.072\\
Street 31 m & 0.9 & 0.352 & 0.291 & 0.208 & 0.185 & 0.078\\
 & 1.2 & 0.498 & 0.336 & 0.347 & 0.233 & 0.126\\
 & 1.5 & 0.552 & 0.296 & 0.407 & 0.182 & 0.070\\
Rooftop 97 m & 0.9 & 0.063 & 0.043 & 0.015 & 0.014 & 0.003\\
 & 1.2 & 0.177 & 0.453 & 0.062 & 0.054 & 0.036\\
 & 1.5 & 0.260 & 0.534 & 0.124 & 0.113 & 0.095\\\bottomrule
\end{tabular}
\end{table}

Table~\ref{tab:veh} is a descriptive sweep over all three sites. At the street-level site reversals are common, the full spectrum recovers 17 to 37 points, and at $1.2\lambda$ and $1.5\lambda$ the quadratic rule is 11 to 23 points above Spec. At the rooftop sites reversals are rare at $0.9\lambda$, and at $1.2\lambda$ and above the azimuth-only geometry rule is worse than the nearest port, which motivated the use of the elevation in the line-of-sight rule. A separate rooftop run at 108~m with 1500 queries at $0.5\lambda$, $0.7\lambda$ and $0.9\lambda$ gave nearest-port reversal rates of 0.003, 0.003 and 0.026 against an oracle of 0.

\emph{Confirmation and reproduction.} Three predictions for the street-level site were fixed before a confirmation run with a new seed and 2000 queries. The nearest port is at least 10 points above Spec, Spec is at least 5 points above Orc, and at $1.2\lambda$ the quadratic rule is at least 5 points above Spec. The run gave the nearest port, the azimuth rule, Spec, the quadratic rule and Orc rates of 0.356, 0.299, 0.192, 0.212 and 0.080 at $0.9\lambda$ and 0.475, 0.334, 0.221, 0.326 and 0.124 at $1.2\lambda$, and all three predictions held. An independent reproduction with 500 queries and per-query outputs gave paired 95 percent bootstrap intervals for the nearest port minus Spec of 16.4 $[12.8,20.2]$ and 24.6 $[20.0,29.0]$ points, for Spec minus Orc of 9.0 $[6.6,11.6]$ and 10.2 $[7.6,13.0]$ points, and for the quadratic rule minus Spec of 1.2 $[-0.4,3.0]$ and 7.8 $[4.6,11.2]$ points at $0.9\lambda$ and $1.2\lambda$. The insufficiency of the second moment is therefore significant at $1.2\lambda$ only.

\section{Unsurveyed Positions: Full Results}\label{app:real3}
Table~\ref{tab:real3full} lists all map rules at the street-level site for five survey densities. The survey of density $g$~m keeps survey points at least $g$~m apart, chosen greedily in a random order, and the sparser surveys are subsets of the denser ones. The selected rows report the rule chosen on the validation set, among the four maps without history and, for the history option, also VS-Hist. The learned predictors were not trained for the 20~m survey of 13 points.

\begin{table*}[t]\centering
\caption{Reversal rate at unsurveyed positions, Chicago street-level site, $d=0.9\lambda$, 2400 test queries in 22 cells, by survey density (number of survey points in parentheses). The rule selected on the validation set is named in brackets.}\label{tab:real3full}
\begin{tabular}{lccccc}\toprule
Rule & 20 m (13) & 10 m (31) & 5 m (79) & 2 m (236) & Full (4517)\\\midrule
Nearest & 0.494 & 0.494 & 0.494 & 0.494 & 0.494\\
LoS with elevation & 0.366 & 0.366 & 0.366 & 0.366 & 0.366\\
IDW, score $c$ & 0.397 & 0.324 & 0.315 & 0.340 & 0.328\\
IDW, score~\eqref{eq:risk} & 0.395 & 0.321 & 0.316 & 0.330 & 0.319\\
VS, score $c$ & 0.346 & 0.330 & 0.286 & 0.281 & 0.316\\
VS, score~\eqref{eq:risk} & 0.345 & 0.306 & 0.290 & 0.280 & 0.313\\
VS-Hist & 0.275 & 0.265 & 0.254 & 0.248 & 0.270\\
Selected without history & 0.346 [VS] & 0.330 [VS] & 0.315 [IDW] & 0.281 [VS] & 0.319 [IDW]\\
Selected with history option & 0.275 [VS-Hist] & 0.265 [VS-Hist] & 0.315 [IDW] & 0.248 [VS-Hist] & 0.270 [VS-Hist]\\
Learned without history & -- & 0.318 & 0.287 & 0.282 & 0.318\\
Learned with history & -- & 0.259 & 0.241 & 0.244 & 0.267\\
Spec (reference) & 0.215 & 0.215 & 0.215 & 0.215 & 0.215\\
Score~\eqref{eq:risk} from local paths (reference) & 0.218 & 0.218 & 0.218 & 0.218 & 0.218\\
Orc (reference) & 0.090 & 0.090 & 0.090 & 0.090 & 0.090\\\bottomrule
\end{tabular}
\end{table*}

\begin{table*}[t]\centering
\caption{Reversal rate at unsurveyed positions for further spacings, the scattering case and the rooftop site, full survey unless stated. The rooftop site has 3120 test queries in 19 cells and a full survey of 3057 points, and its 5~m survey has 63 points.}\label{tab:real3more}
\begin{tabular}{lccccccc}\toprule
 & \multicolumn{2}{c}{Street level} & Street, scattering 0.3 & \multicolumn{3}{c}{Rooftop (108 m)}\\
Rule & $0.7\lambda$ & $1.2\lambda$ & $0.9\lambda$ & $0.9\lambda$ & $1.2\lambda$ & $0.9\lambda$, 5 m survey\\\midrule
Nearest & 0.200 & 0.625 & 0.485 & 0.071 & 0.373 & 0.071\\
LoS with elevation & 0.268 & 0.382 & 0.357 & 0.006 & 0.095 & 0.006\\
Selected without history & 0.194 & 0.353 & 0.328 & 0.009 & 0.097 & 0.009\\
VS-Hist & 0.180 & 0.349 & 0.379 & 0.006 & 0.094 & 0.005\\
Learned without history & 0.193 & 0.343 & 0.322 & 0.018 & 0.122 & 0.298\\
Learned with history & 0.173 & 0.339 & 0.319 & 0.030 & 0.141 & 0.376\\
Spec (reference) & 0.150 & 0.324 & 0.216 & 0.004 & 0.095 & 0.004\\
Orc (reference) & 0.019 & 0.190 & 0.085 & 0.002 & 0.089 & 0.002\\\bottomrule
\end{tabular}
\end{table*}

Paired intervals of the selected map without history minus Spec are $[5.1,17.0]$ points with the full survey and $[5.5,19.6]$ with the 10~m survey at $0.9\lambda$, so the map does not reach the local statistics. The nearest port minus Spec is $[23.5,31.3]$ points, and Spec minus Orc is $[8.9,16.9]$ points. At $1.2\lambda$ with the 10~m survey the learned predictor with history beats the selected rule by 1.5 points $[0.6,2.6]$, which is significant but below the threshold of 3 points. With scattering coefficient 0.3, the validation set selected a map without history fit also for the history option at every density except 20~m. With the full and 10~m surveys VS-Hist is indeed worse than the selected map on the test set, which is consistent with transported specular paths that no longer describe the diffuse channel, although we did not isolate this cause. With the 5~m and 2~m surveys VS-Hist would have been better, at 0.280 and 0.249 against 0.315 and 0.284. At the rooftop site the geometry rule is within 0.2 points of Spec at $0.9\lambda$, and the learned predictors are worse than the selected map at every density at which they were trained. At the rooftop site the nearest port reverses 7.1 percent of the returns at $0.9\lambda$ and 37.3 percent at $1.2\lambda$, where the line-of-sight rule and the map rules stay within 1 point of Spec (Table~\ref{tab:real3more}). Choosing by the score~\eqref{eq:risk} instead of $c$ from the local paths changes the reversal rate by less than one point at all spacings and sites.

\section{Five Cities: Full Results}\label{app:cities}
The history-axial rule fits the forward history with orthogonal matching pursuit over a grid of 41 axial direction cosines and delays, with eight atoms, and predicts the candidates from the axial component $\boldsymbol r^{\mathsf T}\hat{\boldsymbol v}$ of the residual only. The street-axis rule applies the line-of-sight score along the longest free direction of the local geometry description. For each held-out city the strongest geometry rule was selected on the validation queries of the training cities, and it was the line-of-sight rule with elevation in all five cases.

\begin{table}[t]\centering
\caption{Zero-survey port selection in five cities, all ten sites pooled, 38184 queries in 3759 clusters. Reversal rate, with the mean clipped alignment $\E[\bar s]$ in parentheses.}\label{tab:xsceneall}
\setlength{\tabcolsep}{2.6pt}
\begin{tabular}{lccc}\toprule
Rule & $0.7\lambda$ & $0.9\lambda$ & $1.2\lambda$\\\midrule
Nearest & 0.126 (0.607) & 0.367 (0.200) & 0.573 ($-0.125$)\\
Street axis & 0.292 (0.350) & 0.479 (0.022) & 0.476 (0.032)\\
History axial & 0.282 (0.367) & 0.487 (0.031) & 0.485 (0.023)\\
LoS with elevation & 0.132 (0.635) & 0.205 (0.496) & 0.285 (0.358)\\
Learned & 0.120 (0.636) & 0.214 (0.461) & 0.305 (0.292)\\
Learned with history & 0.119 (0.630) & 0.225 (0.423) & 0.316 (0.266)\\
Spec (reference) & 0.086 (0.717) & 0.144 (0.603) & 0.251 (0.423)\\
Orc (reference) & 0.014 (0.954) & 0.068 (0.827) & 0.169 (0.664)\\\bottomrule
\end{tabular}
\end{table}

Table~\ref{tab:xsceneall} extends Table~\ref{tab:xscene} to all spacings. The line-of-sight rule minus the learned predictor is 1.1 $[0.4,2.0]$, $-0.9$ $[-1.2,-0.7]$ and $-2.0$ $[-2.4,-1.6]$ points at $0.7\lambda$, $0.9\lambda$ and $1.2\lambda$, and minus the learned predictor with history 1.2 $[0.1,2.6]$, $-2.0$ $[-2.9,-1.1]$ and $-3.1$ $[-3.8,-2.4]$ points. The learned predictor with history stays 3.3 $[2.7,4.2]$, 8.1 $[6.5,9.9]$ and 6.4 $[5.5,7.6]$ points above Spec. Both axis rules are worse than the nearest port at $0.7\lambda$ and $0.9\lambda$. Table~\ref{tab:sites} lists the sites. At $0.7\lambda$ the line-of-sight rule is better than the nearest port at all sites except the two Chicago sites, and at $0.9\lambda$ and $1.2\lambda$ it is better at all ten.

\begin{table*}[t]\centering
\caption{Zero-survey reversal rate per site and spacing. Learned is the predictor without history trained on the other four cities.}\label{tab:sites}
\begin{tabular}{lccccccccccc}\toprule
 & \multicolumn{3}{c}{Nearest} & \multicolumn{3}{c}{LoS with elevation} & \multicolumn{3}{c}{Learned} & Spec & Orc\\
Site & $0.7\lambda$ & $0.9\lambda$ & $1.2\lambda$ & $0.7\lambda$ & $0.9\lambda$ & $1.2\lambda$ & $0.7\lambda$ & $0.9\lambda$ & $1.2\lambda$ & $0.9\lambda$ & $0.9\lambda$\\\midrule
Chicago street & 0.188 & 0.495 & 0.666 & 0.253 & 0.362 & 0.412 & 0.217 & 0.371 & 0.438 & 0.219 & 0.102\\
Etoile street & 0.205 & 0.538 & 0.663 & 0.163 & 0.278 & 0.323 & 0.161 & 0.296 & 0.358 & 0.220 & 0.108\\
Florence street & 0.233 & 0.499 & 0.616 & 0.222 & 0.340 & 0.342 & 0.205 & 0.348 & 0.357 & 0.259 & 0.120\\
Munich street & 0.221 & 0.512 & 0.600 & 0.210 & 0.312 & 0.347 & 0.191 & 0.327 & 0.366 & 0.232 & 0.093\\
San Francisco street & 0.175 & 0.539 & 0.737 & 0.142 & 0.252 & 0.379 & 0.144 & 0.292 & 0.380 & 0.212 & 0.115\\
Chicago rooftop & 0.004 & 0.054 & 0.362 & 0.005 & 0.008 & 0.077 & 0.005 & 0.009 & 0.091 & 0.004 & 0.001\\
Etoile rooftop & 0.123 & 0.518 & 0.733 & 0.104 & 0.204 & 0.340 & 0.101 & 0.223 & 0.375 & 0.166 & 0.101\\
Florence rooftop & 0.124 & 0.393 & 0.588 & 0.099 & 0.172 & 0.329 & 0.098 & 0.181 & 0.344 & 0.139 & 0.057\\
Munich rooftop & 0.120 & 0.427 & 0.605 & 0.092 & 0.158 & 0.307 & 0.096 & 0.169 & 0.338 & 0.131 & 0.067\\
San Francisco rooftop & 0.048 & 0.260 & 0.526 & 0.043 & 0.075 & 0.242 & 0.044 & 0.080 & 0.251 & 0.062 & 0.017\\\bottomrule
\end{tabular}
\end{table*}

\section{Error Budget: All Settings}\label{app:budget}
Table~\ref{tab:budget} lists every setting of Section~\ref{sec:budget}. The history signal-to-noise ratio is relative to the mean power of the history of each query. The phase drift of the $m$-th history sample is a random walk with the given standard deviation per 40~ms step, anchored at the forward time. The relative position error is added independently to each history position and as one common error to the candidate residuals, which represents an error in $\boldsymbol D$, with the given standard deviation per axis. The absolute position error moves the horizontal position used for the survey neighbors, the virtual-source transport and the direction of the base station. The orientation error rotates the body offsets $\boldsymbol p=\boldsymbol r-\boldsymbol D$ about the origin. The calibration error is a fixed phase of each return port relative to the forward port. The moderate joint setting combines 10~dB, $5^\circ$, 2~mm, 1~m, $2^\circ$ and $10^\circ$, and the severe one 5~dB, $15^\circ$, 5~mm, 3~m, $5^\circ$ and $20^\circ$. The criterion that a gain is kept, a lower interval end above zero, was fixed before the run.

\begin{table*}[t]\centering
\caption{Error budget at unsurveyed positions, Chicago street level, $d=0.9\lambda$, full survey, 2400 test queries in 22 cells. Reversal rates, and the gain of VS-Hist in percentage points with paired cluster-bootstrap 95 percent intervals (2000 resamples).}\label{tab:budget}
\setlength{\tabcolsep}{3.5pt}
\begin{tabular}{lccccccccc}\toprule
Setting & Nearest & LoS & Map & VS-Hist & Orc & Nearest $-$ VS-Hist & LoS $-$ VS-Hist & Map $-$ VS-Hist & LoS $-$ Map\\\midrule
No error & 0.494 & 0.366 & 0.319 & 0.270 & 0.090 & 22.4 [15.0, 28.2] & 9.5 [1.3, 20.9] & 4.9 [$-0.3$, 11.2] & 4.7 [$-0.4$, 10.7]\\
SNR 20 dB & 0.494 & 0.366 & 0.319 & 0.269 & 0.090 & 22.5 [15.2, 28.3] & 9.7 [1.4, 21.1] & 5.0 [$-0.2$, 11.5] & 4.7 [$-0.4$, 10.7]\\
SNR 10 dB & 0.494 & 0.366 & 0.319 & 0.276 & 0.090 & 21.8 [14.6, 27.4] & 9.0 [0.9, 20.2] & 4.3 [$-0.8$, 10.5] & 4.7 [$-0.4$, 10.7]\\
SNR 5 dB & 0.494 & 0.366 & 0.319 & 0.301 & 0.090 & 19.3 [12.3, 24.9] & 6.5 [$-1.5$, 17.6] & 1.8 [$-3.3$, 7.9] & 4.7 [$-0.4$, 10.7]\\
Drift $5^\circ$ & 0.494 & 0.366 & 0.319 & 0.278 & 0.090 & 21.6 [14.1, 27.5] & 8.8 [0.5, 20.0] & 4.1 [$-1.2$, 10.5] & 4.7 [$-0.4$, 10.7]\\
Drift $15^\circ$ & 0.494 & 0.366 & 0.319 & 0.296 & 0.090 & 19.8 [13.3, 24.8] & 7.0 [$-1.7$, 18.8] & 2.3 [$-3.2$, 9.1] & 4.7 [$-0.4$, 10.7]\\
Drift $30^\circ$ & 0.494 & 0.366 & 0.319 & 0.345 & 0.090 & 15.0 [9.4, 19.4] & 2.1 [$-6.5$, 14.1] & $-2.5$ [$-8.3$, 4.5] & 4.7 [$-0.4$, 10.7]\\
Relative 2 mm & 0.496 & 0.371 & 0.322 & 0.284 & 0.090 & 21.3 [13.9, 26.8] & 8.8 [0.8, 20.3] & 3.8 [$-1.4$, 10.3] & 5.0 [$-0.3$, 11.2]\\
Relative 5 mm & 0.515 & 0.380 & 0.339 & 0.312 & 0.090 & 20.2 [13.9, 25.5] & 6.8 [$-1.0$, 17.9] & 2.7 [$-1.7$, 7.8] & 4.2 [$-1.7$, 11.2]\\
Relative 1 cm & 0.532 & 0.395 & 0.378 & 0.399 & 0.090 & 13.2 [7.9, 17.6] & $-0.4$ [$-7.2$, 9.1] & $-2.1$ [$-5.7$, 2.3] & 1.8 [$-2.6$, 7.4]\\
Absolute 1 m & 0.494 & 0.365 & 0.316 & 0.279 & 0.090 & 21.5 [14.4, 27.3] & 8.7 [0.5, 20.0] & 3.7 [$-0.9$, 9.5] & 5.0 [$-0.2$, 11.2]\\
Absolute 3 m & 0.494 & 0.370 & 0.336 & 0.316 & 0.090 & 17.8 [10.7, 24.1] & 5.3 [$-1.6$, 15.0] & 2.0 [$-1.8$, 6.3] & 3.4 [$-2.1$, 9.9]\\
Orientation $2^\circ$ & 0.498 & 0.364 & 0.324 & 0.271 & 0.090 & 22.8 [15.4, 28.6] & 9.3 [1.0, 20.6] & 5.3 [0.1, 11.7] & 4.0 [$-1.2$, 10.1]\\
Orientation $5^\circ$ & 0.505 & 0.378 & 0.326 & 0.285 & 0.090 & 22.0 [14.9, 27.6] & 9.3 [1.5, 20.4] & 4.1 [$-1.0$, 10.4] & 5.3 [0.1, 11.1]\\
Orientation $10^\circ$ & 0.509 & 0.395 & 0.370 & 0.321 & 0.090 & 18.8 [12.1, 23.9] & 7.4 [1.1, 16.0] & 4.9 [0.5, 9.7] & 2.5 [$-1.6$, 7.4]\\
Calibration $10^\circ$ & 0.495 & 0.371 & 0.325 & 0.274 & 0.093 & 22.2 [15.1, 27.7] & 9.8 [1.4, 21.2] & 5.1 [$-0.1$, 11.4] & 4.7 [$-0.7$, 11.1]\\
Calibration $20^\circ$ & 0.489 & 0.375 & 0.335 & 0.294 & 0.092 & 19.5 [12.6, 24.9] & 8.1 [0.1, 18.4] & 4.2 [$-1.0$, 9.8] & 3.9 [$-0.8$, 9.6]\\
Calibration $30^\circ$ & 0.495 & 0.393 & 0.349 & 0.305 & 0.085 & 19.0 [13.0, 24.0] & 8.8 [1.6, 18.6] & 4.4 [$-0.3$, 9.8] & 4.4 [0.0, 9.7]\\
Joint moderate & 0.499 & 0.371 & 0.328 & 0.300 & 0.092 & 19.9 [12.9, 25.1] & 7.1 [$-1.2$, 19.0] & 2.8 [$-2.2$, 9.4] & 4.3 [$-1.1$, 10.7]\\
Joint severe & 0.511 & 0.396 & 0.379 & 0.385 & 0.087 & 12.7 [7.8, 16.6] & 1.2 [$-6.0$, 11.4] & $-0.6$ [$-3.8$, 3.4] & 1.8 [$-3.6$, 8.7]\\\bottomrule
\end{tabular}
\end{table*}

In two settings the interval of the gain over the selected map has a lower end just above zero, an orientation error of $2^\circ$ and of $10^\circ$. Without errors this gain is not significant, so we do not read these two settings as a gain that the errors create. The selected map here is the inverse-distance mixture with the risk score that the validation set selected for Table~\ref{tab:real3}, with four neighbors and equal weights.

\section{Training Experiments}\label{app:train}
\emph{Surrogate.} The surrogate trains a ten-class linear softmax classifier on the digits data set for 3000 steps with minibatches of 32, with the features standardized by the statistics of the training split. Each update multiplies the exact minibatch gradient by a value of $\bar s$ drawn from the paired test queries of the rule, and an additive Gaussian gradient noise is the only other noise. The learning rate of each rule is selected on a validation split, and the test accuracy is averaged over five seeds. At $0.9\lambda$ the mean test accuracies are 87.8 percent for the nearest port, 94.9 for the line-of-sight rule, 95.2 for the selected map, 94.9 for VS-Hist, 94.8 for the learned predictor with history, 94.9 for Spec, 94.7 for Orc and 95.1 without aging. At $0.7\lambda$ all rules lie between 94.7 and 95.1 percent. The rules other than the nearest port reached similar final accuracies, and across the seven rules other than the ideal return the Spearman coefficient between accuracy and reversal rate was 0.04 at $0.9\lambda$. On this convex task with tuned learning rates, the rules with a clearly positive mean alignment reached similar accuracies, and the accuracy dropped, with a large spread across seeds, only when the mean alignment approached zero.

\emph{Scalar link.} The experiment of Table~\ref{tab:d1} uses the same digits data set, converted to a binary task of the digits below five against the others, with 1400 training, 200 validation and 197 test samples and standardized features with a bias, and the logistic loss of the equalized output. The channels are normalized by the root mean square of the forward history of each query. Each minibatch of 32 samples draws 32 test queries, and each sample uses the forward channel and the return channel of the port chosen by the rule at its query. The terminal computes $\delta=\mathcal L'(z)/\hat h_f^*$ with the estimated forward channel and returns it only when $|\hat h_f|^2\ge0.1$, which skips 6.5 percent of the samples under ideal estimation. The return noise is set relative to the mean power of the returned $\delta$ over the transmitted samples. The learning rate of each rule is selected from $\{0.01,0.03,0.1,0.3,1\}$ by the mean validation accuracy over three seeds, and ten further seeds give the test accuracy. The exact minibatch gradient averages $\mathcal L'(z)\R(h_f/\hat h_f)\boldsymbol x$ over the samples, which is the derivative of the loss along the forward computation with the estimated channel. The cosine is computed per step and averaged over steps and seeds, and the projection coefficient is the median over the steps of a run, averaged over seeds. At 10~dB the cosines are 0.035, 0.246, 0.336, 0.421, 0.699 and 0.902 in the order of Table~\ref{tab:d1}. The final accuracies of the line-of-sight rule and VS-Hist differ by 1.0 and 1.2 points at 20 and 10~dB, and their intervals overlap. We did not compute paired intervals of the accuracy differences between rules.

\section{Implementation Details}\label{app:impl}
\emph{Data split.} At the street-level site the 9600 queries of the data set are grouped by the 10~m cell of the first step of their trajectory. Thirty percent of the cells, rounded to whole cells, form the test set and 15 percent the validation set, and training and validation queries closer than 2~m to any test query are dropped. The rooftop site is split in the same way.

\emph{Map rules.} For IDW the number of neighbors $K\in\{1,2,4,8\}$ and the exponent $p\in\{0,1,2\}$ are chosen on the validation set for each survey density. VS-Hist uses $K=2$ neighbors and $p=0$ with the full survey of the zero-scattering street-level data, and these two parameters are selected on validation data in every other setting. It uses the $L=24$ strongest transported paths, history positions spaced by 4.8~cm, which is $0.56\lambda$, along the trajectory, 32 subcarriers spaced by 0.625~MHz within a nominal 20~MHz band, and a ridge parameter of $10^{-3}$. The fit forms an $L\times L$ normal matrix from an $MN_f\times L$ design matrix and solves it.

\emph{Learned predictors in Chicago.} The network has two hidden layers of widths 128 and 64 with dropout 0.1, and the history version adds a branch of the same widths for the history. Training uses the binary cross-entropy loss with logits, Adam with a rate of $10^{-3}$, weight decay $10^{-4}$, batches of 256 and at most 60 epochs, and stops after 10 epochs without improvement of the validation reversal rate. A single run with one random seed was made.

\emph{Learned predictors across cities.} These predictors use the same architecture and optimizer with batches of 512, at most 80 epochs and a patience of 12 epochs, without the map scores, which do not exist in an unsurveyed city. They are trained on the queries of four cities and stopped on their validation queries.

\emph{Pedestrian data.} For each query the solver traces the forward position, the 12 candidate positions for the three spacings of the data set, and eight history positions in one call, in groups of 25 queries. The channels are summed over all paths found by the solver, and up to the 160 strongest paths of each query are stored as its path list, which Spec and the map rules use. The available experiment scripts, result summaries, frozen specifications and subsequent amendments are retained together.

\bibliographystyle{IEEEtran}
\bibliography{refs}
\end{document}